# A Vision for Numerical Weather Prediction in 2030

By

Tim Palmer
Department of Physics
University of Oxford

11 June 2020

*In this essay, I outline a personal vision of how I think Numerical Weather Prediction (NWP) should evolve in the years leading up to 2030 and hence what it should look like in 2030. By NWP I mean initial-value predictions from timescales of hours to seasons ahead. Here I want to focus on how NWP can better help save lives from increasingly extreme weather in those parts of the world where society is most vulnerable.*

*Whilst we can rightly be proud of many parts of our NWP heritage, its evolution has been influenced by national or institutional politics as well as by underpinning scientific principles. Sometimes these conflict with each other. It is important to be able to separate these issues when discussing how best meteorological science can serve society in 2030; otherwise any disruptive change - no matter how compelling the scientific case for it - becomes impossibly difficult.*

<u>1. Why Bother?</u>

Why bother? Some might argue that NWP is pretty good already and serves its purpose warning society of weather-related natural hazards. Perhaps no great transformation in NWP capability is needed, just steady, modest, evolution. However, if NWP is so good, why is still frequently the case that when a major weather event hits some part of the world, devastating the local population, it is only days later that emergency relief is distributed to the stricken population? Although we can point to examples where NWP has triggered action ahead of an extreme event, why isn't its use more universal, particularly in the developing world where society is often most vulnerable to these events? Why isn't it being used <u>enough</u>, given its potential to save property and lives?

An important reason why it isn't being used enough was brought home to me quite forcibly at a workshop on ensemble prediction discussing the devastating 2018 Kerala floods. A speaker proudly announced that the ECMWF ensemble prediction system had forecast, a week ahead of time, that a 2-sigma rainfall event would hit Kerala with around 30% probability (apparently with higher probability than any other ensemble forecast system). It was concluded that the ECMWF forecast provided a credible warning of an extreme event. However, one of the Indian scientists at the meeting shook his head in disbelief at the naivety of this assessment: no-one in Kerala would even notice a 2-sigma event, he commented - this was close to a 15-sigma event! Of course, for this type of genuinely extreme event, the probability of occurrence a week ahead was essentially zero for reasons discussed below. The implication, of course, is that a week ahead, the extreme flooding event was not predicted with anything like the level of probability that would make a governmental or aid agency take notice. It would not, for example, trigger the type of forecast-based-finance initiative discussed below. Perhaps two or three days before the event, an inkling of its intensity became apparent. However, at such a short lead time it was much more difficult to get emergency relief in place ahead of time.

It is my strong belief that we need to develop our NWP systems considerably so that by 2030 they become reliable predictors of genuinely extreme events, at least in the range where substantial proactive actions can be planned and executed - say at least 7-10 days ahead of time. Disruptive thinking is needed if we are to achieve this.

We would like our seasonal forecasts to provide credible estimates of the likelihood of occurrence of seasonal extremes. However, on seasonal timescales matters are much worse than in the medium-range – with model biases typically as large as the signals being predicted. For example, current-generation models substantially under-simulate the long-lived quasi-stationary circulation anomalies that generate predictability on the seasonal timescale, largely because of problems with scale interactions as discussed below [1]. As a result, probabilistic reliability on the seasonal timescale is frequently poor [2]. In truth, seasonal predictions are a long way from being genuinely useful. It is hard to imagine that many real-world decisions are strongly influenced by current seasonal forecast systems (except perhaps in countries very close to the El Niño event).

Of course, developments in NWP over the last years – increased model resolution, better data assimilation, improved parametrisations, larger ensembles with stochastic parametrisation – have all improved our ability to forecast extreme events such as tropical and extratropical cyclones, at least in the medium range [3]. Steady developments will see further improvements over the coming years. However, in my view this will leave us a long way short of what could and indeed should be achieved by 2030 across the spectrum of extreme weather events, if only we had the human and computational resources. And yet weather and climate science is not badly funded compared to many other sciences. The real problem is that, in my view, we do not use the funds we are collectively given by the tax payers of the world, efficiently and effectively.

Current weather events are extreme enough. However, most climate-change forecasts suggest that as the atmosphere warms and holds more water vapour, extreme weather events – drought, storm, heat wave - will become yet more extreme. That is to say, improving early-warning systems on multiple timescales are crucial in helping society become more resilient to the changing nature of weather and climate extremes. Hence, we should therefore see improving our NWP systems as a vitally important part of the climate adaptation programme. Indeed, according to the Global Commission for Adaptation [4], improved early-warning systems has the greatest benefit/cost ratio of any climate adaptation measure.

## 2. Scientific Principles

Whatever the NWP system of 2030, it must be firmly founded on scientific principles. This is not the "motherhood and apple-pie" statement that it might appear to be, because, as discussed below, in current NWP practice, meteorological science has been compromised as a result of national or institutional politics. This has led to the development of NWP systems where science is not optimally benefitting society.

One of the great strengths of meteorology is that our predictions are falsifiable, which as Karl Popper emphasised many years ago is a necessary condition for any subject to be considered scientific. Of course, everyone agrees with the importance of falsifiability, but what do we conclude when a deterministic weather prediction is falsified - when reality turns out differently to what was predicted? The traditional conclusion has been that the forecast model or initial condition need to be improved. But this is not the correct conclusion. In a chaotic system like climate, single deterministic forecasts can be expected to be falsified all the time, no matter how good the model and initial conditions are. Indeed, when the atmosphere is in a particularly unstable part of the climate attractor, a forecast can be falsified dramatically – by failing to predict or misplacing an extreme event completely – even when the model and initial conditions are relatively accurate. What then is the conclusion? The conclusion is that the philosophy of determinism is at fault. This is not a matter of forecast timescale. Even at the very shortest forecast ranges (think about convective storm systems), forecasts can be sensitive to errors in model equations or observational input.

Hence, if we aspire to be scientific, we should not be providing single deterministic forecasts to users or to the public. However, it is still the case that most NMSs (and even international forecast organisations) continue to disseminate so-called high-resolution single deterministic forecasts (from either a global or limited-area model). The argument is sometimes made that this is what the customer wants. However, the customer also wants forecasts to be reliable. A reliable deterministic forecast (no matter how high the resolution of the model) is simply a contradiction, a non-sequitur: it is unscientific. In any case, as discussed below, there is now a much better way to provide high-resolution detail within an ensemble context. As such, the practice of disseminating deterministic forecasts (including, incidentally, ensemble-mean forecasts which are also deeply flawed as predictions of the real world since they necessarily damp partially predictable extreme events) should cease by 2030 – if not before! Instead, all predictions should be ensemble based [5], and of course the probabilities produced by these ensembles should be reliable. It is of unrealistic to expect the raw ensemble data to be fully reliable, and calibration corrections must be made to ensure statistical reliability.

There is a second important point to be made about falsification. It is vital if forecast users are to assess the reliability of products derived from global ensemble systems, that sufficient hindcast data is made available. This is a highly nontrivial and computationally intensive requirement. It means that every time the forecast system changes, a new set of hindcasts (or "reforecasts") must be made stretching back over many years to provide adequate sample sizes. Reforecast data is also needed to provide the calibrations needed to make the probabilistic forecasts reliable, and also to train the AI-based systems for downscaling (see below). Not many centres have the resources to provide such ensemble reforecast data sets. However, again, they are essential if we aspire to be scientific. Many

NMSs produce limited-area model forecasts with either no or little reforecast data and no interface to users where they can interrogate the skill of the model. This is both unscientific and unhelpful for users.

## 3. High-Resolution Global Ensembles

The single most important development in NWP in the coming years will be the implementation of global ensemble prediction systems with grid-point spacing (or spectral equivalent) of 1-3km, where some of the key atmosphere/ocean processes (deep convection, orographic gravity wave drag and ocean mesoscale eddies) are represented by the laws of physics rather by approximate semi-empirical parametrisation formulae as is presently the case. There are three reasons for suggesting this:

i) Extreme events are often associated with coherent nonlinear organised or self-aggregated structures, such as associated with mesoscale convection (leading to extreme precipitation), or with persistent quasi-stationary anticyclones (leading to extreme drought). There is ample evidence that high-resolution models capture better the scale interactions that are crucial for maintaining these types of extreme events. Conversely, there is little doubt that parametrisations can do damage to these scale interactions, e.g. by diffusing away sharp gradients which are in reality being maintained by processes like nonlinear wave breaking. The existence of fat tails in climatic distributions [6] provides evidence that these extreme events have predictability. However, the failure to simulate these tails in current-generation models frequently makes them inadequate tools for many extreme-event predictions (like the Kerala floods).

ii) In current-generation models, we are unable to assimilate the crucial information in our observational network due to inadequate model resolution. For example, the skill of medium-range forecasts has been shown to be relatively poor when upstream initial conditions are dominated by organised mesoscale convective systems [7]. In this case, there is no shortage of observations of such convective systems. However, because the assimilating model cannot resolve such systems, the information in the observations is either aliased, damped, or simply wrongly assimilated into the model, leading to poor forecasts downstream, days later.

This is a very important point often not fully understood. It is typically assumed that the best way to improve the initial conditions of a weather forecast is to increase the number of high-quality observations. In the cases mentioned in the paragraph above, the forecasts would not have improved much no matter how many extra observations became available. Increasing the resolution of the assimilating model would instead have been a much more effective way of improving these initial conditions. In such circumstances, the benefit/cost ratio of improving the initial conditions by increasing the assimilating model resolution to O(1-3km) would likely be much greater than improving initial conditions by funding a new satellite observing system. The budgets of the space agencies are measured in the billions per year. And yet the value of this investment in terms of improved forecast skill is being diminished because of insufficient investment in modelling capability. Investment in the development of high-resolution global data assimilation could be transformative for improving initial-condition accuracy and hence forecast skill. There is an imbalance here.

iii) The third reason why improving resolution is so important is that it will help reduce the systematic errors in model which, as mentioned, are typically of the same magnitude as the signals such models attempt to predict [8]. In current generation models, large-scale systematic errors (e.g. in the persistence of coherent quasi-stationary circulations) start to become significant after about two weeks into a forecast, and of course they persist into the climate-change timescales, making regional estimates of climate change extremely uncertain.

The development of 1-3km global ensemble NWP should become possible by 2030 using Tier-0 exascale supercomputing and where numerical precision is reduced to a level consistent with the stochasticity in the system (certainly 32-bit floating-point reals, and likely 16-bit floating-point reals for many parts of the model). The use of AI, trained on existing parametrisations but replacing existing parametrisations can reduce the computational cost of the parametrisations. Domain-specific languages can help make the dynamical cores more computationally efficient, especially on mixed CPU/GPU computing platforms.

Having said this, I believe it is important not to compromise ensemble size for increasing model resolution. 50-member ensembles are important for two reasons: to provide enough probabilistic resolution to forecast extreme events, and to provide enough probabilistic resolution to forecast multivariate events. (As a simple example of the latter, consider forecasts of electricity production from renewable energy sources. To make such forecasts one needs to know the correlated probabilities that both the sun is shining and the wind is blowing. These correlated probabilities cannot be estimated reliably with small ensemble sizes.) NWP can certainly benefit from larger ensemble sizes, but 50 should be seen as a minimum size.

### 4. Downscaling and Calibrating to Postcode Scale: AI rather than LAMs.

As mentioned, it is common for NMSs to run their own limited-area model system, embedded in a global forecast system, in order to provide locally enhanced resolution. Of course, these models simply inherit the systematic biases of the global models at their lateral boundaries. As also mentioned, most NMSs do not have the resources to run a full 50-member ensemble of limited-area model integrations and they also do not have the resources to produce the reforecasts necessary to validate the forecast system adequately, and develop and test products derived from the limited-area model ensembles. They therefore fail in some of the key criteria needed if we aspire to develop a scientific NWP system by 2030.

Fortunately, there is an alternative to limited-area modelling and one that I am personally much more enthusiastic about. In recent years, there have been considerable advances in the use of AI (e.g. with Generative Adversarial Networks [9]) to do high-resolution stochastic downscaling. My own view is that instead of running small limited-area model ensembles, we should instead be applying these sophisticated AI schemes (trained using the reforecast data from the global ensemble) to directly calibrate and downscale the output from global ensembles to postcode scales. Scientists from NMSs should be trained in these AI schemes, applying them to the postcode scales in their own countries and using national observations to optimise the calibrations.

The fact that AI based systems are already competitive with very short-range limited-area model forecasts [10] provides further evidence that limited-area models may not have a useful role for NWP in 2030. My own conclusion, therefore, is that limited-area models should be phased out in the coming years and should likely play no operational role by 2030. NMS computing resources saved could be usefully spent producing the reforecasts for the global ensemble system. There is clearly an important role for limited area modelling for weather and climate research purposes, but not in my view for operational NWP. It is time to rethink this area of operational forecasting.

One can legitimately ask whether AI can completely replace NWP altogether. For medium-range prediction I think this is extremely unlikely. An ability to forecast reliably the probability of highly nonlinear phenomena in the medium range (e.g. second-generation cyclones) using NWP (i.e. using the laws of physics), requires high-quality models run from high-quality initial conditions. To do this with the same level of skill using AI would likely require an exceptional (and hence unrealistic) amount of training data. How much training data do you need to provide adequate data analogues for AI-based medium-range prediction? It has been estimated [11] that it would take about $10^{30}$ years of data collection before we can find two data analogues of the atmosphere over the Northern Hemisphere which agree to within observational error! This same point can also be seen in a different way. If we look at adjoint sensitivity vectors [12], which provide the patterns of initial conditions needed to optimally improve specific forecasts, they are dominated by flow-dependent sub-synoptic-scale features which then evolve into synoptic and larger scales. This upscale evolution is a consequence of non-self adjointness which itself is a consequence of the linearised form of the nonlinear advection term in the equations of motion. The small-scale features in the sensitivity vectors themselves lie well down the spectrum of scales provided by an Empirical Orthogonal Function (EOF) analysis. High-resolution NWP can resolve such sensitivity vectors and one can say that this is a reason why increasing the resolution of NWP models has been a successful strategy over the years. However, a data-driven scheme, trained on limited past data, will only perform well when the information which drives the scheme lies in the space of leading EOF modes. It will not work well when the crucial information is buried deep (and therefore difficult to access) in the sup-space of modes which, from a "variance-explained" perspective, are deemed totally unimportant. For these reasons, I do not believe AI will ever compete seriously with NWP in the all-important day 7-10 medium range.

By contrast, on seasonal timescales AI is already starting to become competitive with NWP [13]. However, this likely reflects the fact that seasonal NWP forecasts are so degraded by model systematic errors that it is relatively easy to be competitive with them. The answer here, in my opinion, is not to give up doing seasonal prediction by NWP, but instead to invest in the models so that these systematic errors are substantially reduced. We need good physics-based models in any case to study climate change. Here the methods of AI completely fail (extrapolation is their Achilles Heel, and they simply cannot be used to extrapolate so far into an unprecedented future). The benefits of a good seamless NWP/climate change prediction system cannot be underestimated [14].

### 5. Impacts, Forecast-based-Finance and Public-Private Partnerships: COP26

Stochastic downscaling and calibration are vital for running impact models e.g. hydrology, agronomy, health, energy and so on. To be specific, consider a potentially very important application of NWP in 2030: for determining when an agency might send forecast-based-finance [15] to a developing-world country at risk for some particular severe weather event (flood, drought, wind storm etc). An analysis will have been done ahead of time to determine a suitable trigger for sending such forecast-based-finance. Typically, this will be based on a prior-determined probabilistic threshold: when the probability of some particular extreme event (potentially defined from multiple meteorological variables or from a user-impact model driven by such variables) exceeds this threshold, then the country will receive the funds e.g. to help buy and locate emergency supplies ahead of time. Again, a week's lead time would seem to be the shortest feasible if such a scheme is to be effective.

For such forecast-based-finance schemes to be viable, the forecast probabilities must be properly calibrated and be fully reliable. Otherwise the agencies will either too-often fail to send funds ahead of especially devastating events and the schemes will be written off as ineffective, or conversely limited funds will be frittered away by acting in vain with large numbers of so-called false alarms (leading donors to withdraw funding). This is why having considerable past reforecast data is so vital, and why basing triggers on poorly validated limited-area model output is so totally unsatisfactory.

A crucial question is how data from global ensemble systems can be relayed to NMSs, especially developing country NMSs. It does not seem feasible to transmit raw data – there is too much of it. Instead, much of the downscaling and calibration should either be done in-house by the global forecast centres, or it should be done at specialist regional centres such as RIMES. Given that such downscaling and calibration will help developing countries improve their early-warning systems, and given that this should legitimately be considered part of the climate adaptation agenda for the developing world, such downscaling and calibration work could be done under the auspices of (e.g. Copernicus) Climate Services.

The private sector (especially large companies like Google, IBM, Microsoft etc) have considerable expertise with AI and they can certainly contribute to providing the software needed to calibrate and downscale the global ensemble data for the developing country NMSs and also doing AI-based rapid-update nowcasting. However, in providing calibrated downscaled forecast data to aid agencies and developing-country NMSs, the software should be provided free of charge. I do not think that the use of NWP systems developed by the private sector is itself a viable option for use by developing country NMSs and aid agencies: these are not global ensemble systems and they do not have adequate hindcast datasets to provide the required validation.

However, so that developing country NMSs can optimally benefit from global ensemble systems, it is necessary that the global ensemble forecast centres provide all their raw data to aid agencies and developing-country NMSs free of charge. I believe that a unanimous gesture in this direction should be made by the global forecast centres (and the private companies who may contribute through AI for example) in time for COP26 in 2021. As mentioned, improving our early-warning systems is a legitimate part of the climate adaptation programme.

As mentioned, the most important key to progress lies in providing the global ensemble centres with Tier 0 supercomputing and adequate human resources. Of course, supercomputing is provided by the private sector, though the funding comes from the tax payer. The extent to which new types of private-partnerships need to be developed is unclear to me, though philanthropic donation from the private sector for Tier-0 supercomputing for extreme-event prediction should certainly be encouraged. What would be much more transformational than new private-public partnerships would be the reorganisation of the NMSs so that they work together cooperatively in helping to develop the global ensemble forecast systems, instead of pursuing their own fragmented limited-area model work. If scientists at the NMSs could work collaboratively towards the development of very small numbers (see below) of global high-resolution ensemble systems, the consequences could be revolutionary!

### 6. A collaborative NWP development programme

It is important that scientists from developing countries can also contribute to the development of the global ensemble system – it will become divisive if NWP somehow becomes solely a "first-world" activity. Of course, some operational centres are mindful of this today and do take part in development programmes and such like. However, my sense is that the potential for developing country scientists to contribute to the science of NWP is much greater than is currently being realised. For example, open-source versions of operational ensemble systems must be made available to run, e.g. at regional centres such as RIMES, themselves with Tier-1 supercomputing. When a global ensemble system does not perform well on a specific extreme event, regional scientists need to be able to study the factors that the forecast was sensitive to, and make recommendations on how to improve the global forecast system. Forums need to be further developed where these insights can be fed back to the primary forecast centres and implemented in new operational cycles. NWP needs the scientific insights from scientists around the world. Centres like ICTP in Trieste should be developed to encourage interaction between NWP (and climate) scientists from developed and developing countries.

How many global ensemble models do we need? In answering this question, it is vital not to think of a "model" as a deterministic piece of code. If models were deterministic, then indeed we might need many models in order to span model uncertainty. However, by 2030 all models (weather and climate) should explicitly encode the uncertainties introduced by the numerical truncation of the underlying partial differential equations, through explicit stochasticity [16]. The scientists who develop models must recognise model uncertainty as a primitive concept in model development, and not a tag-on extra. For example, if there is no unambiguous closure scheme for the parametrisation of a physical process (e.g. cloud microphysics), then parametrisation schemes which stochastically combine multiple possible closures should be developed (model uncertainty is certainly not just about parameter perturbations). If stochastic models are developed which encompass the uncertainty in closure schemes, there is then no compelling case for maintaining and developing multiple models, especially since individual institutes do not in any case have the resources to maintain and develop high-resolution ensemble forecast systems, particularly given the needs for the extensive reforecasts discussed above.

My own opinion is that one (stochastic) model system per continent is about right. By pooling human and computational resources across the NMSs and academic sectors within each continent, and focussing on a very small number of modelling systems, it should be possible by 2030 for NWP ensemble forecast systems with the required 0(1-3km) to be run out to the seasonal timescale, and supplemented with sophisticated AI for stochastic downscaling and calibration (and rapid-update nowcasting) made possible by plentiful reforecast data with the same forecast system. The calibrated data would then be fed into a range of impact models producing probabilistic forecasts of electricity production, river discharge, flood risk, storm surge, structural damage, health impact, crop yield and so on. This latter activity, I would suggest, should be the primary mandate of NMSs in 2030 (see below). By contrast, if resources are split between multiple centres within a continent, as is currently the case, there is a significant risk that we will not get to this happy state by 2030 and society will not be optimally served.

I am fully aware that some NMSs may cite national security as a reason for requiring a semi-independent NWP system. However, I believe there are ways to overcome such concerns (e.g. by having independent copies of internationally developed code which could potentially be run at lower resolution on some national Tier 1 computing resource in times where national security issues demand it). However, as mentioned at the beginning, it is important that we carefully separate scientific from such other issues in our discussions about the future of NWP.

## 7. The NMS mandate and the role of the forecaster in 2030

I finish with some remarks about what I see should be the primary mandate of the vast majority of NMSs in 2030. My view is that it should simply be that of turning forecast data from the global ensembles into useful products for the various national impact sectors. As mentioned, many NMSs do not have the capacity to receive the global ensemble data, so much of the analysis would need to be done by regional centres such as RIMES, or in-house at the ensemble forecast centre. As also discussed above, this will require performing regionally-specific calibration of the global ensemble data against observations. Particularly useful in this respect will be the specific national weather observations that the NMSs are mandated to make – here is where these observations can have particular value (perhaps more so than assimilating them into the global forecast system). Rapid update short-range predictions can be made blending AI-based schemes, using the latest regional observations, with the latest global ensemble output. As mentioned above, I see no compelling requirement for limited-area NWP in this scheme of things.

What about the role of forecasters in 2030? As Allan Murphy [17] wrote: "Forecasts possess no intrinsic value. They acquire value through their ability to influence the decisions made by users of the forecasts". The role of the human "forecaster" in 2030 will be to help the user make the right decisions using available probability forecasts. Like the wavefunction of quantum theory under measurement, this is where – and not before - probabilities finally "collapse" into deterministic actions and decisions. Here, prior to the forecast itself, the meteorologist will have interacted with the user on how to determine the probability threshold for a relevant type of weather event (which may be a compound event based on multiple meteorological variables), above which certain actions should be taken (e.g. evacuating a neighbourhood, or moving valuable livestock to higher ground, given a possible storm surge). This is not at all straightforward and not a process I see AI doing any time soon: it is likely to require prolonged in-depth interaction with the user. However, through such interaction, when a user needs to make a decision in real time, the decision process should then be relatively straightforward and would simply be a matter of assessing whether the pre-determined threshold has indeed been exceeded or not. This is already starting to happen now. However, by 2030, this type of procedure should be commonplace around the world and in the developing world in particular. The silly notion that "the public will never understand the notion of probability" (an old ruse by an NMS for not accepting probabilistic over deterministic predictions) will be finally consigned to the history books. In this way, the human meteorologist (what we would now call a "forecaster" although in 2030 he or she may not actually be doing much forecasting *per se*) will certainly have an extremely important role in the NWP system of 2030.

If this can be achieved, I believe we will no longer be asking, in 2030, why isn't NWP being used <u>enough</u>? Rather early-warning will be the ubiquitous go-to tool to help society become more resilient to the ever-greater intensity of weather extremes.

**Acknowledgements**

My thanks to Erin Coughlan, Peter Webster and Antje Weisheimer for helpful comments on an early draft of this paper. This work was supported by the European Research Council Advanced Grant ITHACA, 741112.